\documentclass[a4paper,onecolumn,nopdfoutputerror,11pt]{quantumarticle}
\pdfoutput=1
\usepackage[utf8]{inputenc}
\usepackage[english]{babel}
\usepackage[T1]{fontenc}
\usepackage{amsmath}
\usepackage{hyperref}
\usepackage{tikz}
\usepackage{lipsum}

\usepackage{graphicx}%
\usepackage{multirow}%
\usepackage{amssymb,amsfonts}%
\usepackage{amsthm}%
\usepackage{mathrsfs}%
\usepackage[title]{appendix}%
\usepackage{xcolor}%
\usepackage{textcomp}%
\usepackage{manyfoot}%
\usepackage{algorithm}%
\usepackage{algorithmicx}%
\usepackage{algpseudocode}%

\usepackage{subcaption}
\usepackage{dcolumn}
\usepackage{bm}

\theoremstyle{thmstyleone}%
\newtheorem{Theorem}{Theorem}
\newtheorem{Proposition}[Theorem]{Proposition}
\newtheorem{Corollary}[Theorem]{Corollary}

\begin{document}

\title{Vacuum-Signal Detection and the Principle-Level Feasibility of Arbitrarily Long-Distance Repeaterless Quantum Communication}

\author{Hao Shu}
\affiliation{Sun Yat-sen University}
\orcid{0000-0002-9332-7103}
\email{Hao\_B\_Shu@163.com}

\maketitle

\begin{abstract}
  Practical repeaterless quantum communication (PRQC) is constrained by the divergence of quantum bit error rate (QBER) arising from the interplay between channel loss and single-photon detector (SPD) dark counts. As the channel transmission rate decays with distance, vacuum-signal-induced dark counts inevitably dominate detection events beyond a finite range, driving QBER toward 50\% and rendering PRQC infeasible. Here, a theoretical framework termed vacuum-signal detection (VSD) is established to address this limitation at the level of principle. By employing controlled operations together with multi-copy analysis, the VSD paradigm enables vacuum-induced detection events to be identified and filtered without disturbing the encoded messages. Consequently, the non-vacuum signal ratio (NVSR) of the accepted signals can be stabilized at a high value irrespective of channel attenuation, thereby suppressing the fundamental QBER of PRQC within the secure bounds at any communication distance. As a result, PRQC can, in principle, remain feasible over arbitrarily long distances. By providing a rigorous theoretical resolution of the fundamental QBER-induced distance limitation, this work clarifies the principle-level scalability of PRQC. Furthermore, the vacuum-filtering framework developed here and its introduction of multi-copy analysis may also be of interest in a broader class of loss-sensitive or detection-based quantum tasks.
\end{abstract}


\section{Introduction}

Practical repeaterless quantum communication (PRQC), most prominently exemplified by quantum key distribution (QKD) \cite{BB1984Quantum,LY2018Overcoming}, enables information exchange with information-theoretic security beyond classical cryptographic limits \cite{S1994Algorithms}. Its security is rooted in fundamental quantum-mechanical principles, such as the no-cloning theorem, and is realized through single-photon-level signal preparation, error correction, and privacy amplification \cite{GL2004Security}. Over the past decades, PRQC has advanced from foundational theoretical proposals to sophisticated implementations, with demonstrations extending from controlled laboratory environments to long-haul intercity links\cite{LC2012Measurement,LY2018Overcoming,LZ2023Experimental,LW2025Entanglement}.

A fundamental requirement for PRQC is that the quantum bit error rate (QBER) remains below a protocol-dependent threshold. In practice, however, the interplay between channel attenuation and single-photon detector (SPD) dark counts inevitably degrades the signal-to-noise ratio, manifesting as a fundamental QBER. As transmission loss grows with distance, the non-vacuum signal ratio (NVSR), and thus the effective signal ratio (ESR), at the receiver decreases exponentially, whereas SPD dark counts probabilistically generate random detection reports for ineffective signals with an error probability of $50\%$. Therefore, once the ESR drops to a level comparable to the SPD dark count rate (DCR), the QBER asymptotically approaches $50\%$, eventually surpassing the security thresholds and rendering PRQC theoretically infeasible. Since the security of PRQC is based on the no-cloning theorem, it is widely believed that this QBER explosion cannot be mitigated by classical methods such as multi-copy analysis, and constitutes a long-standing fundamental bottleneck to PRQC, even at the level of principle.

Extensive research has focused on mitigating these constraints by optimizing either the ESR or the performance of SPDs. On the physical layer, advances in optical fiber technology have reduced channel attenuation to below $0.15\ \mathrm{dB/km}$ \cite{TS2017Lowest,KY1986Transmission,NK2002Ultra,HH2013Record,MR2015Record}, while free-space links may exhibit even lower losses under favorable conditions. At the protocol level, twin-field QKD and related schemes \cite{LY2018Overcoming,WY2018Twin,MZ2018Phase,ZZ2022Mode} improve the scaling of the ESR from $O(t)$ to $O(\sqrt{t})$, where $t$ denotes the channel transmission rate (which is also the NVSR in these protocols). In parallel, superconducting SPDs have achieved detection efficiencies exceeding $90\%$ with DCRs below $10^{-5}$ \cite{HL2020Detecting,CL2021Detecting,XZ2021Superconducting,CK2023High,S2023Solve,CB2024Single,SW2025Current}. These advances have enabled secure PRQC over distances beyond $1000\ \mathrm{km}$ \cite{LJ2021Field,ZL2023Experimental}. Nevertheless, these advancements do not fundamentally alter the mechanism underlying QBER explosion: the exponential decay of the NVSR (and thus ESR) with distance inevitably leads to a regime where dark counts dominate effective detection signals, driving the QBER toward the $50\%$. To date, a rigorous theoretical framework demonstrating the feasibility of PRQC over arbitrarily long distances, even at the level of principle, remains absent, leaving the ultimate scalability of PRQC conceptually an open problem.

In this work, a theoretical framework for filtering vacuum-induced errors in PRQC is established, hereafter referred to as the vacuum-signal detection (VSD) paradigm. This paradigm integrates auxiliary degrees of freedom (DOF), controlled operations, projective measurements, and multi-copy analysis to distinguish vacuum-originated events from non-vacuum signals. By selectively discarding detection events identified as vacuum-originated, the VSD module can theoretically stabilize the NVSR (and thus ESR) of the retained signals at a high value at any channel transmission rate, thereby suppressing the resulting QBER within an acceptable range at any communication distance. As a consequence, PRQC is shown to remain feasible in principle over arbitrarily long distances, only subject to an efficiency trade-off. The main observation of the proposed method is that, on one hand, the no-cloning theorem in PRQC only promises that the encoding message can not be cloned, but it does not imply that the photon existence information (PEI) can not be cloned either; On the other hand, it is the existence of photons that determines whether a signal is effective, and PEI can be decoupled with the cryptographic message. This work focuses on the formal derivation of the VSD framework and the characterization of its theoretical performance bounds from a general perspective, leaving the practical implementation and integration into specific PRQC to future investigation.

\section{\label{Results}Results}

In this section, the vacuum-signal detection (VSD) paradigm and its theoretical analysis are presented. They establish a rigorous framework for addressing the fundamental QBER explosion problem in PRQC at the level of principle. The architecture of the VSD module is first described, followed by a formal characterization of its capability through an ability theorem, and finally, its implications for PRQC are discussed.

\subsection{\label{VSD}Vacuum-Signal Detection Module}

Let $\rho_{C}$ denote the quantum state encoded in the message DOF $C$ of a PRQC protocol. For definiteness, qubit states $\rho_{C}\in\mathbb{H}_{C}$ with computational basis $\{|0\rangle,|1\rangle\}$ are considered. An auxiliary DOF $L$, associated with the same particle but independent of $C$, is introduced to host the PEI. In addition, $n$ auxiliary particles with DOFs denoted by $L_{i}$ ($i=1,2,\ldots,n$) are employed. A controlled-$F$ operation, denoted by $C(F)$, applies the operation $F$ to the target system when the control is in state $|1\rangle$, leaves the target unchanged when the control is in state $|0\rangle$, leaves the control unaffected in both cases, and extends linearly to superposition states.

\begin{figure}[htbp]
\centering
\includegraphics[width=\linewidth]{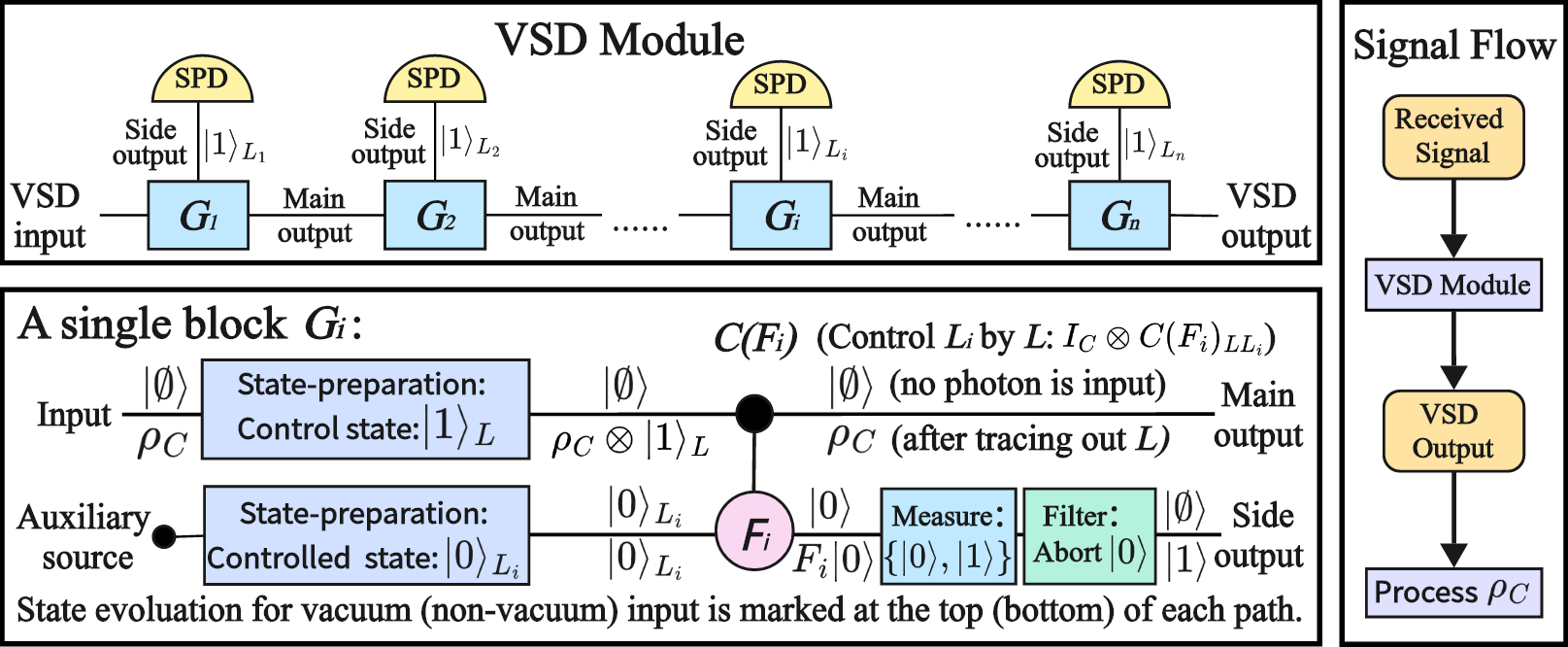}
\caption{
\textbf{Architecture of the VSD module (left) and receiver scheme (right).}
Each incoming signal is first processed by the VSD module, where the auxiliary DOF $L$ of the input particle is coupled to $n$ auxiliary systems $L_{i}$ via controlled operations. The side outputs are then measured projectively, and the number of effective detections in state $|1\rangle$ determines whether the received signal is identified as vacuum or not. Here, $|\emptyset\rangle$ represents that the signal is vacuum.}
\label{ESDM}
\end{figure}

The signal flow of PRQC augmented with the VSD module is illustrated in Fig.~\ref{ESDM}. Each received particle is first processed by the VSD module and then forwarded to the standard message-decoding stage. The VSD module consists of a sequence of $n$ control blocks $G_{i}$ as well as auxiliary projective measurements and SPD detections, implemented as follows:

\noindent (i) The auxiliary DOF $L$ of the incoming particle is prepared in state $|1\rangle$, and each auxiliary particle $L_{i}$ is prepared in state $|0\rangle$.

\noindent (ii) The DOF $L$ is coupled to $L_{i}$ via a controlled operation $C(F_{i})$, with the particle with DOF $L$ acting as the control and the auxiliary particle with DOF $L_{i}$ as the target.

\noindent (iii) Each $L_{i}$ is measured projectively in the $\{|0\rangle,|1\rangle\}$ basis, and only outcomes corresponding to $|1\rangle$ are registered.

\noindent (iv) The control particle is forwarded to the message-decoding stage, while the auxiliary particles are detected by SPDs.

\noindent (v) The number of registered $|1\rangle$ outcomes in the side-outputs is compared with a preset threshold $k$. The ordinary signal is identified as non-vacuum if at least $k$ detection events occur among the $n$ auxiliary channels; otherwise, it is classified as vacuum.

Since the VSD module operates exclusively on the auxiliary DOF of the ordinary particle $L$ that are separable from the message-encoding DOF $C$, and auxiliary particles with DOF $L_{i}$, the encoded quantum information is not disturbed, in principle. The controlled operations allow PEI to be extracted without directly measuring the message DOF or destroying the ordinary particle, while the use of multiple auxiliary blocks enables statistical discrimination that is robust at the level of effective parameters. By circumventing the no-cloning theorem that prevents cryptographic messages from being copied, and copying the fixed (known) PEI of the signals, VSD enables multi-copy analysis and decision in quantum communication to enhance the signal-to-noise ratio as in classical communication. 

To provide physical intuition for the VSD framework, an idealized scenario may be considered in which $L$ and $L_{i}$ correspond to photonic polarization DOFs and $C(F_{i})$ is realized as a controlled-NOT (C-NOT) operation. If the input signal is vacuum, no photon, and hence no polarization DOF, is present on the main path. As a result, neither the preparation of $L$ nor the controlled operations take effect, and the auxiliary photons, initially prepared in state $|0\rangle$, remain unchanged after measurement, which are filtered, yielding no detection events in the side channels. By contrast, for a non-vacuum input, the polarization DOF $L$ is prepared in state $|1\rangle$ and acts as the control of the C-NOT operations, flipping the auxiliary photons from $|0\rangle$ to $|1\rangle$. These auxiliary photons are not filtered and can then be detected by SPDs after the projective measurements. In this way, the presence of the signal is mapped onto the detection statistics of the auxiliary channels, providing a theoretical mechanism to identify and filter vacuum events. This intuition is not restricted to the idealized setting, specific gates, or particular physical DOFs, and is formalized by the ability theorem in Section ~\ref{Ability}, with rigorous proofs given in Section ~\ref{Method}.

\subsection{Notations and Assumptions}

The following notations are adopted throughout this work.

The channel transmission rate of the PRQC is denoted by $t$. When the input to block $G_{j}$ is non-vacuum, $P_{j}$ and $p_{j}$ denote the probabilities of obtaining non-vacuum side and main outputs, respectively, before detection. When the input is vacuum, $Q_{j}$ denotes the probability of obtaining a non-vacuum side output, which characterizes errors associated with state preparation and measurement (SPAM) in the auxiliary path. For simplicity, all blocks $G_{j}$ are assumed to be identical, such that $P_{j}=P$, $Q_{j}=Q$, and $p_{j}=p$ for all $j$.

Let $P_{s}$ and $Q_{s}$ denote the probabilities of registering a detection event in a side output when the input of the corresponding block $G_{j}$ is non-vacuum and vacuum, respectively. Denote by $\eta_{0}$ and $d_{0}$ the detection efficiency and DCR of the SPDs used in the side channels. One then has
\begin{equation}\begin{aligned}
P_{s}&=P[\eta_{0}+(1-\eta_{0})d_{0}]+(1-P)d_{0}
      =P\eta_{0}(1-d_{0})+d_{0},
\end{aligned}\end{equation}
\begin{equation}\begin{aligned}
Q_{s}&=Q[\eta_{0}+(1-\eta_{0})d_{0}]+(1-Q)d_{0}
      =Q\eta_{0}(1-d_{0})+d_{0}.
\end{aligned}\end{equation}

Let $e_{C}$ denote the error rate induced by channel noise and $e_{th}$ the error-tolerance threshold of a PRQC protocol, with $0\leq e_{C}<e_{th}\leq 50\%$. Denote by $\eta_{1}$ and $d_{1}$ the detection efficiency and DCR of the SPDs used in the message-decoding channel. Define
\begin{equation}\begin{aligned}
\gamma:=\frac{(1-2e_{th})d_{1}}{\eta_{1}[e_{th}-e_{C}+d_{1}(1-2e_{th})]}.
\end{aligned}\end{equation}
The following proposition characterizes the minimal tolerable transmission rate of PRQC.

\begin{Proposition}
\label{QCThr}
For a standard PRQC protocol without the VSD module, the condition QBER $< e_{th}$ holds if and only if the ESR exceeds $\gamma$. The parameter $\gamma$ thus establishes a fundamental lower bound on the ESR required to maintain the security of a standard PRQC protocol.
\end{Proposition}

Unless otherwise specified, the following assumptions are adopted. First, the controlled operations in the VSD module are assumed to be in a deterministic manner, and alternative configurations, such as probabilistic controlled operations, can be incorporated within the same framework through appropriate modifications\footnote{For example, in some cases, one can replacing $P_{s}$ by the success rate (after considering SPD) of the controlled operation, namely, the probability of detecting an success controlled operation event, and additionally, replacing the decision criterion from obtaining at least $k$ detection events among $n$ auxiliary channels by that success event rate reaches a pre-defined threshold.}. Second, all device parameters are treated as effective parameters determined by the overall system configuration, independent of the specific choice of engineering realization. For instance, $\eta_{0}$ denotes the effective detection efficiency of the side-channel SPD under the employed auxiliary sources and detection setup, and $P_{s}$ represents the probability of obtaining a detection event in the side channel when the input of block $G_{j}$ is non-vacuum under the corresponding controlled operations. Adopting such effective global parameters, as opposed to device-specific microscopic models, ensures the generality of the theoretical framework and is agnostic to the underlying physical platform.

\subsection{\label{Ability}VSD Ability Theorem}

The central analytical result of the VSD framework is summarized in the following theorem and corollary. Their proofs are deferred to Section~\ref{Method}.

\begin{Theorem}
\label{Thm}
If $pP_{s}>Q_{s}$, then for any channel transmission rate $t$, the NVSR can asymptotically approach
\begin{equation}\begin{aligned}
\label{ThmCod}
\frac{pP_{s}-Q_{s}}{P_{s}-Q_{s}}
\end{aligned}\end{equation}
through appropriate choices of $n$ and $k$ in the VSD module.\footnote{This value also represents the supremum of the NVSR achievable under the VSD architecture described in Section~\ref{VSD}.}
\end{Theorem}

In typical operating regimes, $Q_{s}\ll 1$, as it accounts only for error contributions arising from SPAM errors and SPD DCRs, whereas $pP_{s}$ remains finite, being determined primarily by the controlled operation and the detection efficiency of the side-channel SPDs. As a result, the expression in Eq.~(\ref{ThmCod}) satisfies $\frac{pP_{s}-Q_{s}}{P_{s}-Q_{s}}\approx p$. Given that $p$ represents the intrinsic propagation probability of the main path in a single, non-cascaded controlled block $G_{j}$, and is typically close to unity, the asymptotic NVSR is stabilized at a high constant value, effectively decoupled from the channel attenuation $t$.

The condition $pP_{s}>Q_{s}$ can be rewritten equivalently as
\begin{equation}\begin{aligned}
\label{Cod}
pP>(1-p)\frac{d_{0}}{\eta_{0}(1-d_{0})}+Q.
\end{aligned}\end{equation}
The left-hand side of this inequality is measured by the effects of the controlled operation $C(F)$ (e.g., the quality of a C-NOT gate), while the right-hand side is dominated by noise contributions associated with SPAM and SPD DCRs. Under typical conditions, $\frac{d_{0}}{\eta_{0}}\ll Q\ll 1$, such that the requirement reduces to demanding that the controlled operation be sufficiently reliable compared to the SPAM error present in the absence of $C(F)$. This represents a mild and conceptually natural condition within practically implementable controlled operations. It is important to emphasize that $p$ and $P$ in Eq. \eqref{Cod} are parameters for a single non-cascaded controlled module $G_{j}$ and are not decreased as the number of cascaded modules increases.

The essential implication of Theorem~\ref{Thm} is that the exponentially decaying NVSR encountered in standard PRQC can be converted into a fixed, high, and distance-independent value, as the bound in Eq.~(\ref{ThmCod}) does not depend on the channel transmission rate $t$. Consequently, the fundamental QBER explosion induced by channel loss can be circumvented, suggesting that PRQC can, in principle, be scalable to arbitrarily long distances.

\begin{Corollary}
\label{Cor1}
Assume $pP>(1-p)\frac{d_{0}}{\eta_{0}(1-d_{0})}+Q$, then for any distance (or equivalently channel transmission rate), a PRQC protocol can remain feasible using suitable VSD modules if\footnote{The condition is for protocols whose NVSR equals ESR, for example, BB84 QKD, and is also expected to be necessary. Detailed proof refers to Section \ref{Method}. For advanced protocols in which ESR$>$NVSR, for example, Twin-field QKD, the requirement can be relaxed, with the calculations still following the method presented in the proof of this corollary, only needing a slight modification.}
\begin{equation}\begin{aligned}
\label{FC}
P(p-\gamma)>(1-p)\frac{d_{0}}{\eta_{0}(1-d_{0})}+(1-\gamma)Q.
\end{aligned}\end{equation}
\end{Corollary}

Under typical conditions that $\gamma\ll p$, $d_{0},\gamma\ll 1$, and $\frac{d_{0}}{\eta_{0}}\ll Q$, the corollary effectively reduces to the requirement $pP>Q$, namely the quantity $pP$ determined by the controlled operation exceeds the SPAM error. This criterion is readily satisfied at the level of principle for a controlled operation that is not dominated by noise.

\subsection{\label{Experiment}Numerical Simulations}

\begin{figure}[htbp]
  \centering
  \begin{subfigure}[htbp]{0.48\columnwidth}
    \centering
    \includegraphics[width=\linewidth]{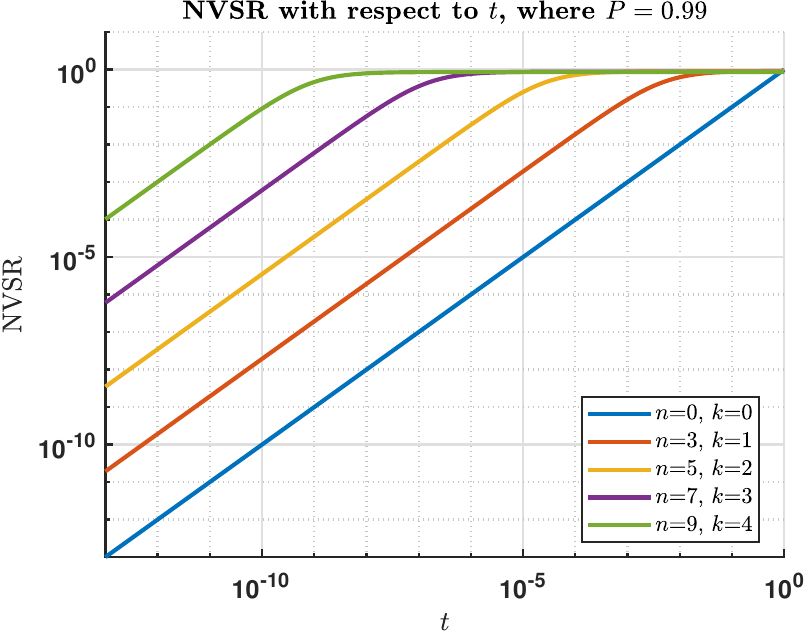}
    \caption{\textbf{NVSR vs $t$, $P=0.99$.}}
    \label{NVSR0.99}
  \end{subfigure}
  \hfill
  \begin{subfigure}[htbp]{0.48\columnwidth}
    \centering
    \includegraphics[width=\linewidth]{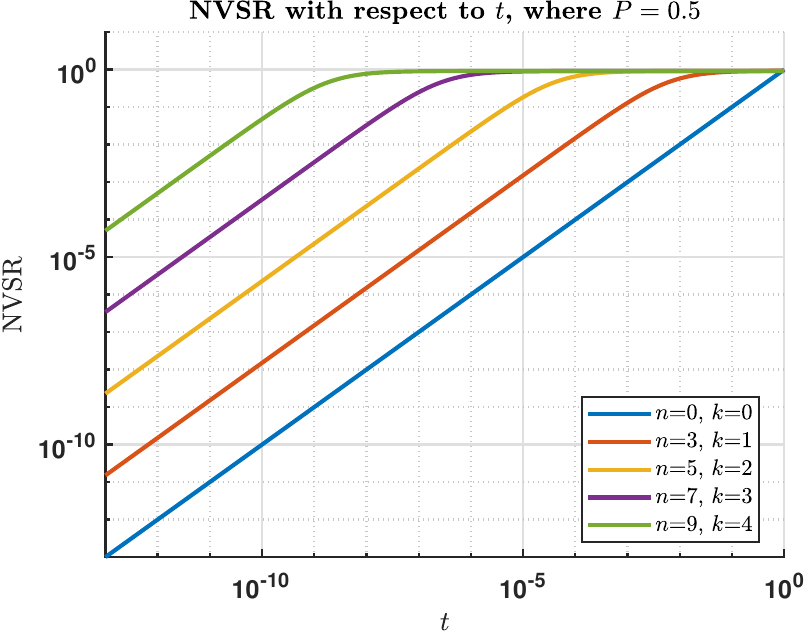}
    \caption{\textbf{NVSR vs $t$, $P=0.5$.}}
    \label{NVSR0.5}
  \end{subfigure}
  \hfill
  \begin{subfigure}[htbp]{0.48\columnwidth}
    \centering
    \includegraphics[width=\linewidth]{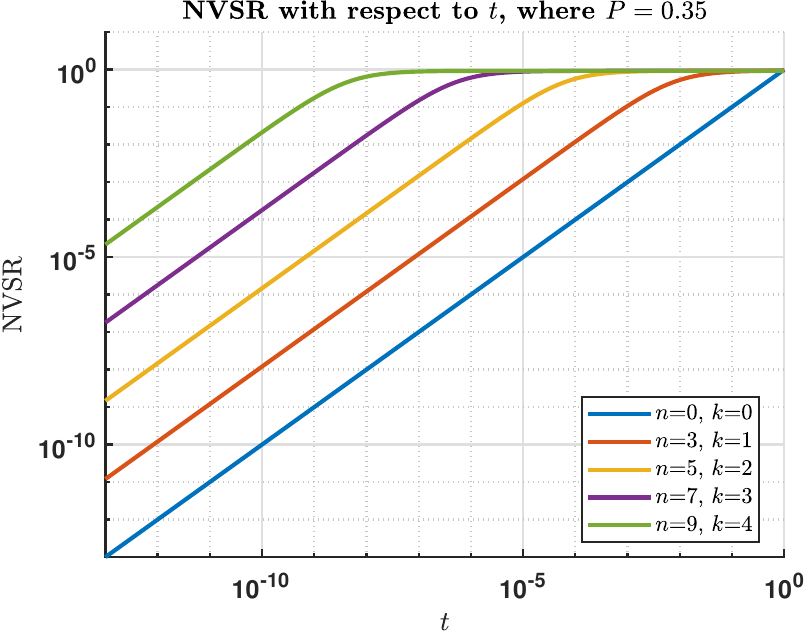}
    \caption{\textbf{NVSR vs $t$, $P=0.35$.}}
    \label{NVSR0.35}
  \end{subfigure}
  \hfill
  \begin{subfigure}[htbp]{0.48\columnwidth}
    \centering
    \includegraphics[width=\linewidth]{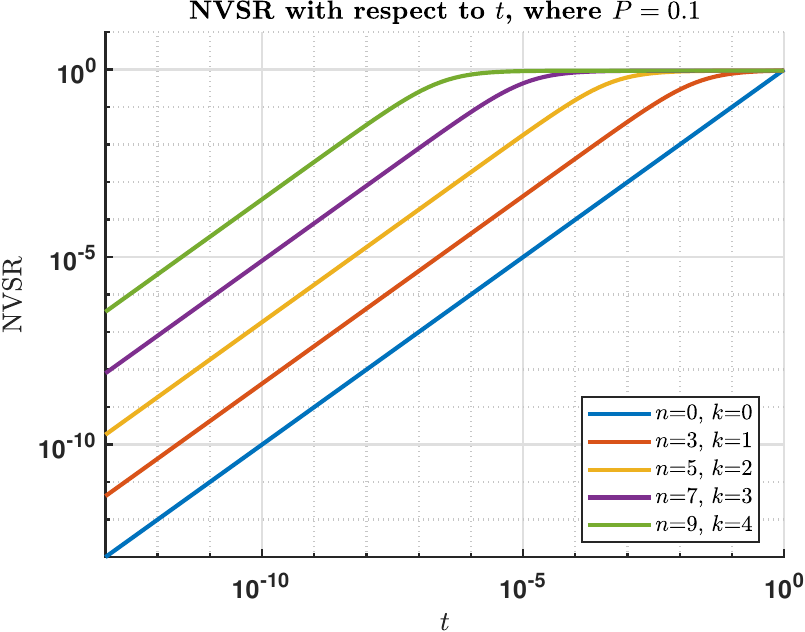}
    \caption{\textbf{NVSR vs $t$, $P=0.1$.}}
    \label{NVSR0.1}
  \end{subfigure}
  \caption{\textbf{NVSR vs $t$ for different controlled-gate parameters.}
The case $n=0,k=0$ corresponds to standard PRQC without VSD. The incorporation of VSD modules leads to a substantial enhancement of the NVSR, especially in the deep-loss (low-transmission) regime, effectively counteracting the exponential decay. For example, with $(n,k)=(7,3)$, the NVSR increases by approximately seven orders of magnitude at $t=10^{-12}$ when $P=0.99$, and by about five orders of magnitude even for a very low-fidelity gate with $P=0.1$. These results validate that the VSD framework effectively stabilizes the NVSR against channel loss.}
  \label{NVSRvst}
\end{figure}

\begin{figure}[htbp]
  \centering
  \begin{subfigure}[htbp]{0.48\columnwidth}
    \centering
    \includegraphics[width=\linewidth]{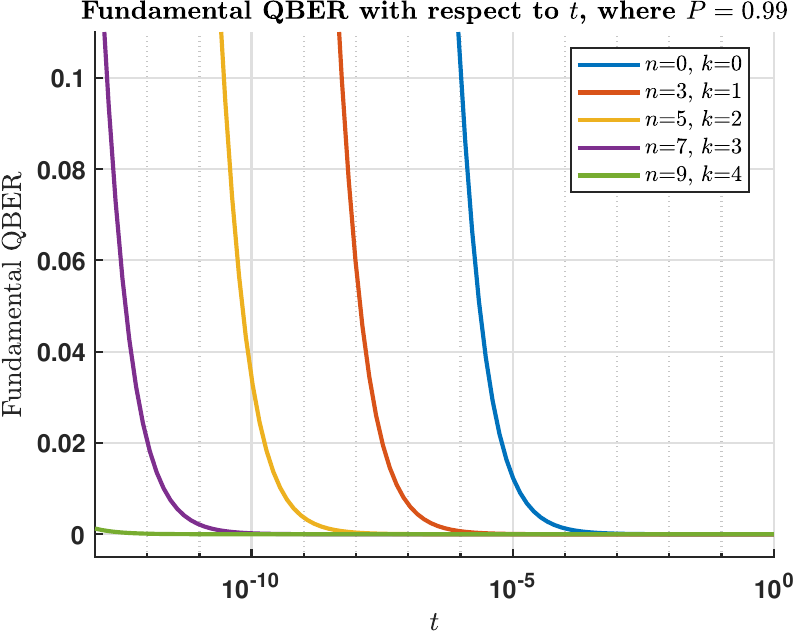}
    \caption{\textbf{QBER vs $t$, $P=0.99$.}}
    \label{QBER0.99}
  \end{subfigure}
  \hfill
  \begin{subfigure}[htbp]{0.48\columnwidth}
    \centering
    \includegraphics[width=\linewidth]{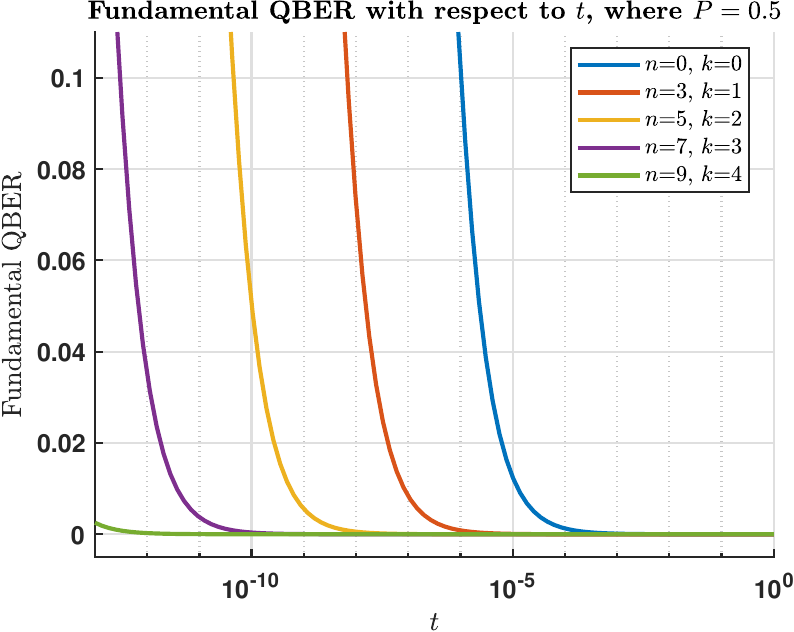}
    \caption{\textbf{QBER vs $t$, $P=0.5$.}}
    \label{QBER0.5}
  \end{subfigure}
  \hfill
  \begin{subfigure}[htbp]{0.48\columnwidth}
    \centering
    \includegraphics[width=\linewidth]{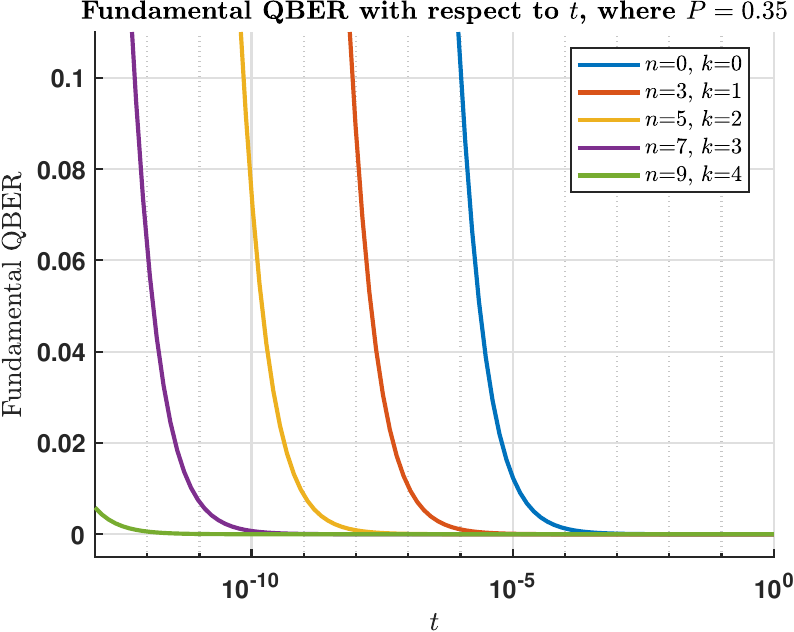}
    \caption{\textbf{QBER vs $t$, $P=0.35$.}}
    \label{QBER0.35}
  \end{subfigure}
  \hfill
  \begin{subfigure}[htbp]{0.48\columnwidth}
    \centering
    \includegraphics[width=\linewidth]{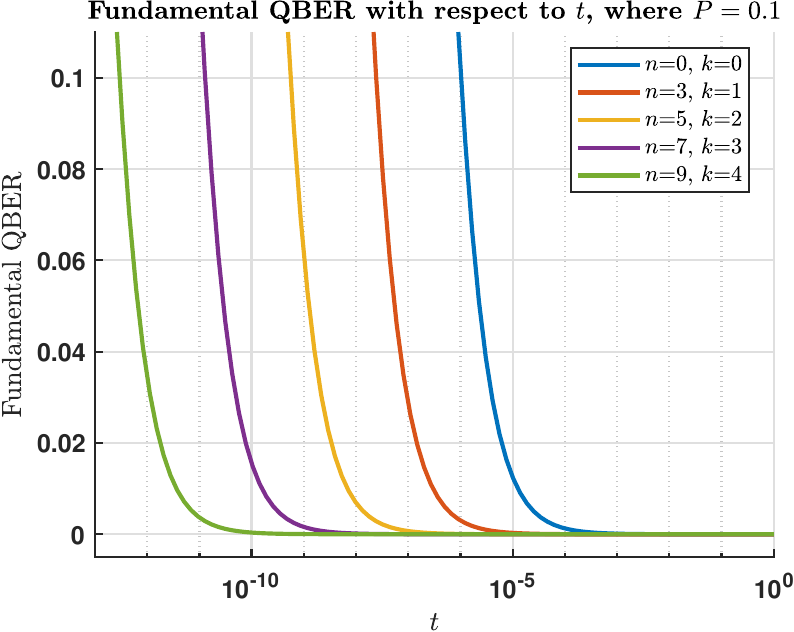}
    \caption{\textbf{QBER vs $t$, $P=0.1$.}}
    \label{QBER0.1}
  \end{subfigure}
  \caption{\textbf{Fundamental QBER vs $t$ under different VSD configurations.}
Here, the simulations are for the case that ESR=NVSR, referring to Section \ref{Method} for details, and for the case that ESR$>$NVSR, the calculation is similar (with only a slight modifications). The application of VSD enables a stable and low QBER even when the transmission rate decreases by several orders of magnitude.
For $(n,k)=(9,4)$, the QBER remains below $3\%$ at $t=10^{-12}$ for both $P=0.99$ and $P=0.1$, whereas without VSD, secure communication ceases once $t\lesssim10^{-6}$. These results underscore the robustness of the VSD paradigm: the QBER remains suppressed even when using gates with limited fidelity, provided the VSD parameters $n$ and $k$ are appropriately tuned.}
  \label{QBERvst}
\end{figure}

\begin{figure}[htbp]
  \centering
  \begin{subfigure}[htbp]{0.48\columnwidth}
    \centering
    \includegraphics[width=\linewidth]{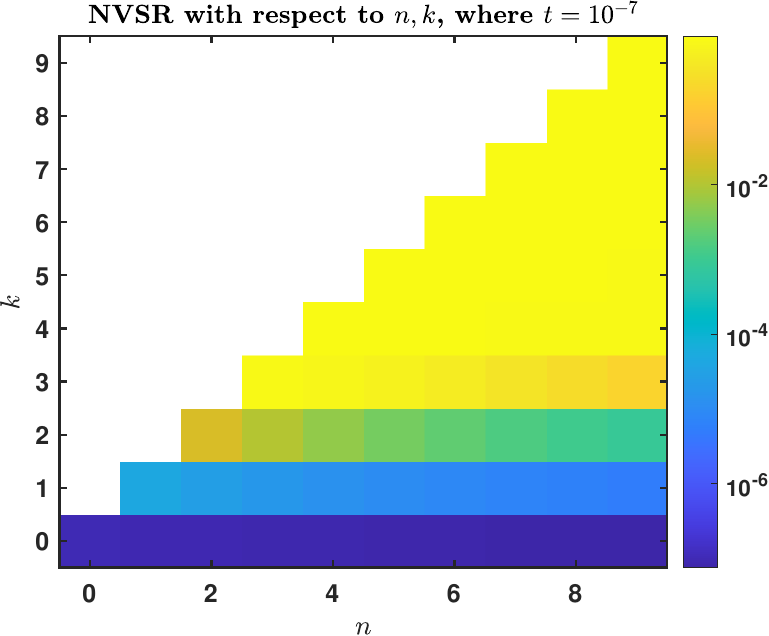}
    \caption{\textbf{NVSR vs $n,k$, $t=10^{-7}$.}}
    \label{FigNVSR-7}
  \end{subfigure}
  \hfill
  \begin{subfigure}[htbp]{0.48\columnwidth}
    \centering
    \includegraphics[width=\linewidth]{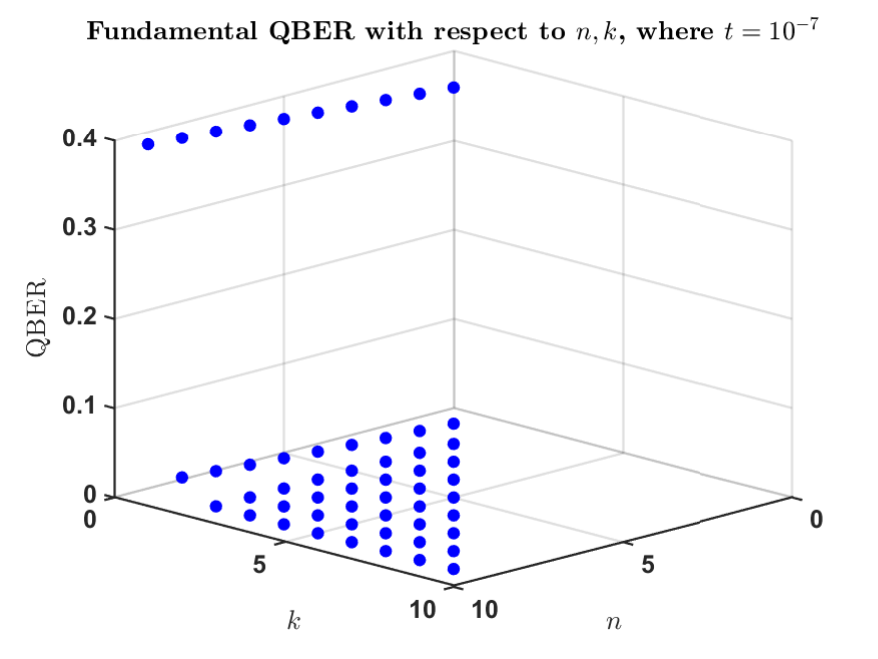}
    \caption{\textbf{QBER vs $n,k$, $t=10^{-7}$.}}
    \label{FigQBER-7}
  \end{subfigure}
  \hfill
  \begin{subfigure}[htbp]{0.48\columnwidth}
    \centering
    \includegraphics[width=\linewidth]{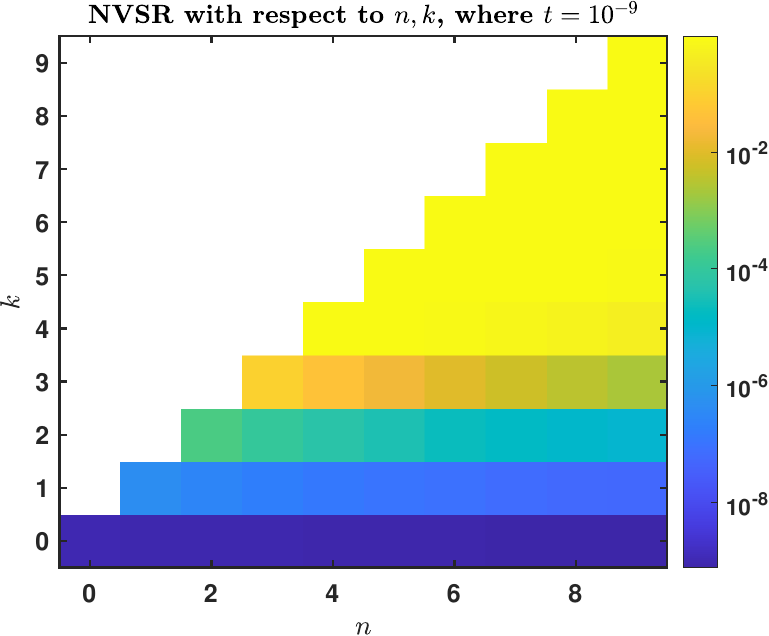}
    \caption{\textbf{NVSR vs $n,k$, $t=10^{-9}$.}}
    \label{FigNVSR-9}
  \end{subfigure}
  \hfill
  \begin{subfigure}[htbp]{0.48\columnwidth}
    \centering
    \includegraphics[width=\linewidth]{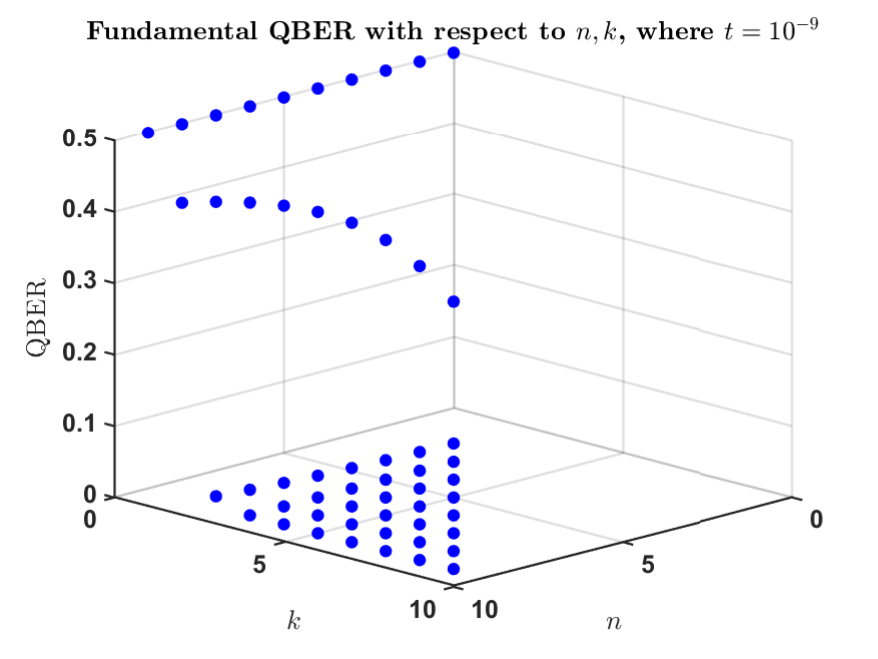}
    \caption{\textbf{QBER vs $n,k$, $t=10^{-9}$.}}
    \label{FigQBER-9}
  \end{subfigure}
  \caption{\textbf{NVSR (Figs.~\ref{FigNVSR-7}, \ref{FigNVSR-9}) and QBER (Figs.~\ref{FigQBER-7}, \ref{FigQBER-9}) as functions of $n$ and $k$ at fixed transmission rates.}
For $t=10^{-7}$, selecting $k\geq 2$ yields a high NVSR, while QBER suppression occurs whenever $k>0$.
For $t=10^{-9}$, $k\geq 3$ ensures a high NVSR, whereas effective QBER suppression occurs whenever $k>1$.
Here $0\leq k\leq n\leq 9$ and $P=0.99$.
These trends suggest that even modest VSD configurations can significantly enhance the NVSR and reduce the QBER.}
  \label{t79}
\end{figure}

To corroborate the analytical results, numerical simulations are performed, as shown in Figs.~\ref{NVSRvst}, \ref{QBERvst}, and \ref{t79}. The simulation results exhibit consistent agreement with our theoretical analysis, demonstrating that even modest VSD configurations are sufficient to suppress the fundamental QBER and significantly relax the minimal tolerable transmission rate $t$ required for PRQC.

\textbf{Simulation settings:} The SPD detection efficiencies are set to $\eta_{0}=\eta_{1}=0.78$, with dark count rates $d_{0}=d_{1}=10^{-7}$ \cite{CK2023High}. The SPAM error is taken as $Q=0.2\%$\cite{HA2014High,P2025A}, and the intrinsic message error is set to $e_{C}=0$, thereby isolating the influence of the fundamental QBER for clearer characterization. The controlled operation $C(F)$ is modeled as a C-NOT gate with $p=0.97$\cite{P2025A}, while the parameter $P$ is varied. Further discussions on representative physical parameter choices are provided in Section~\ref{Device}.

\section{Discussion}

We now discuss the technological feasibility, security implications, efficiency trade-offs, potential circuit-level improvements, and future directions of the proposed VSD framework.

\subsection{\label{Device} Technological Feasibility}

Although the main contribution of this work lies in establishing the theoretical feasibility of the VSD framework that resolves the PRQC problem, it is also important to examine the associated device requirements to assess its relevance for practical implementations. While an experimental realization of the VSD framework would require further systemic integration of quantum technologies, the requisite performance parameters are already within the reach of state-of-the-art optical components.

\begin{figure}[htbp]
\centering
\includegraphics[width=0.6\linewidth]{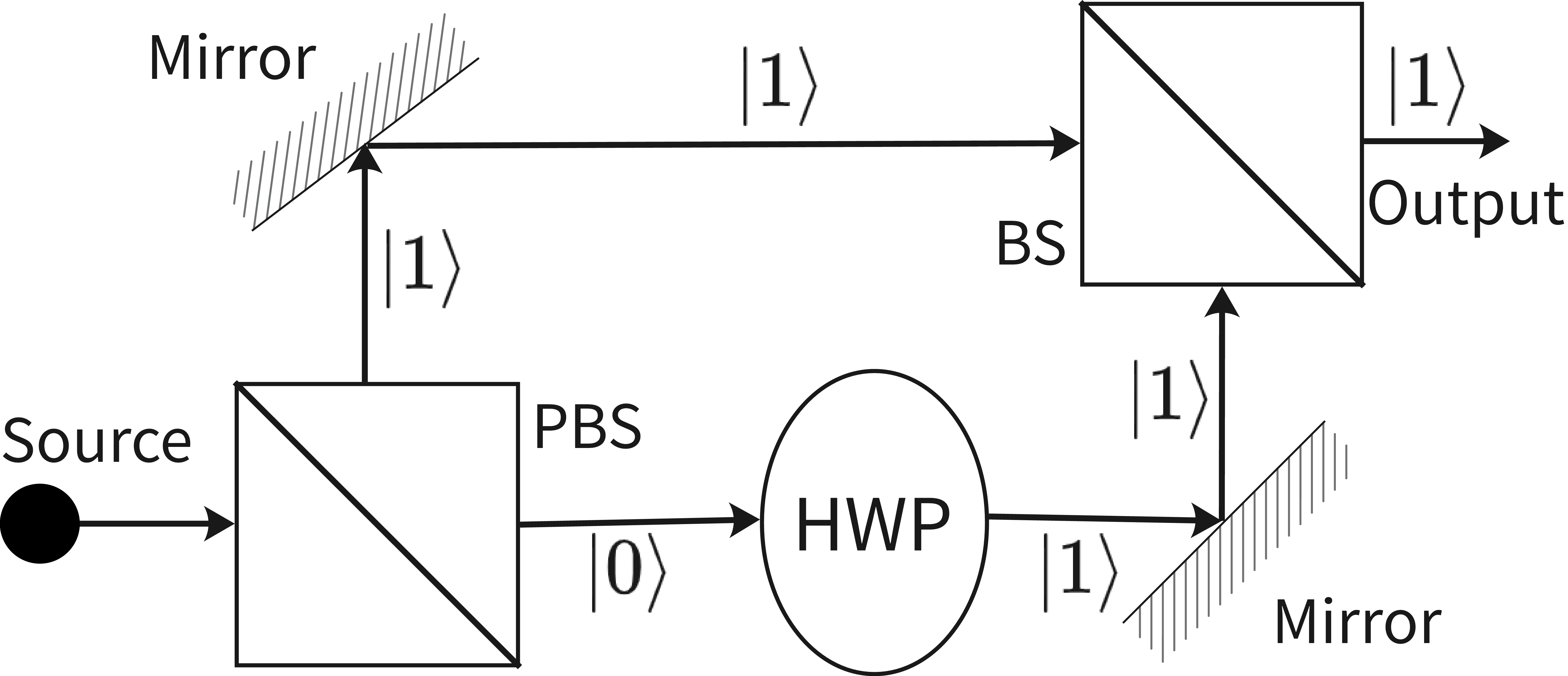}
\caption{\textbf{State preparation for the polarization DOF.} 
A PBS separates the input signal by polarization; one path passes through an HWP that flips the polarization, and the two paths are recombined at a BS, yielding an output with fixed polarization.}
\label{SP}
\end{figure}

In practice, the SPAM error is typically low even for commercial devices, such that $Q\ll1$ is a reasonable assumption \cite{HA2014High,P2025A}. The state preparation and measurement steps of VSD can be implemented using standard linear-optical elements. For example, when the auxiliary DOF $L$ is encoded in photon polarization, state preparation requires only a polarization beam splitter (PBS), a half-wave plate (HWP), and a beam splitter (BS), as illustrated in Fig.~\ref{SP}, while measurement and filtering can be realized using a single polarizer. Under such configurations, preparation error rates below $10^{-3}$ and measurement error rates below $10^{-4}$ have been reported for commercially available photonic components.

The primary technical requirement of the VSD module is the accuracy of the controlled gate $C(F)$, typically C-NOT. Crucially, as per Corollary~\ref{Cor1}, the scalability of PRQC to arbitrarily long distances is guaranteed, provided that the condition in Eq.~(\ref{Cod}) holds. Under typical operating conditions, this reduces to the transparent requirement
\begin{equation}\begin{aligned}
\label{SimpCod}
pP>Q.
\end{aligned}\end{equation}
For C-NOT gates, $P$ represents the correct output probability of the target state, while $p$ corresponds to the survival probability of the control state, independent of its logical correctness. In low-attenuation optical platforms, survival probabilities close to unity are consistent with reported propagation characteristics, for example, attenuation on the order of $0.1,\mathrm{dB}$ (corresponding to $p\approx0.98$), even with solely room-temperature operations \cite{MD2005Propagation,CL2022Ultra,FZ2022Transverse,FZ2025Chip,P2025A}. Moreover, controlled gates with high effective fidelities have been reported under a variety of optical configurations, including feedforward implementations with fidelities exceeding $97\%$ \cite{KH2021Room,HE2023Efficient,SD2025Experimental,AD2025A} and post-selected schemes exceeding $99\%$ \cite{LP2010High,CR2011Integrated,PR2025High,NF2025Quantum,GN2025Quantum,P2025A}. It is important to note that $p$ and $P$ refer solely to the probabilities for a single, non-cascaded $C(F)$ operation and do not decrease with the number of cascaded blocks $G_{j}$, and the accumulation effect has been accounted for in the derivation. Taken together, these observations indicate that the parameter regime required by Theorem~\ref{Thm} and Eq.~(\ref{Cod}) is compatible with existing device-level performance, as it only requires $pP$ to exceed the SPAM error level.

Furthermore, since the auxiliary DOFs employed in VSD are independent of the message-encoding DOF, they can be publicly accessible without compromising security. Consequently, ideal single-photon auxiliary sources are not required to satisfy the security conditions of PRQC protocols. In addition, because the analysis throughout this work is carried out at a global, effective-parameter level rather than relying on device-specific microscopic models, the VSD module does not depend on idealized auxiliary sources in principle, provided that the corresponding parameter constraints are met. Auxiliary preparation using standard optical sources can be compatible with the VSD framework since the requirement of Theorem \ref{Thm} is mild. Intuitively, multi-photon sources might even provide benefits since filters are applied, although further investigations of this are needed. Also, note that nearly perfect sources have been reported even at room-temperature~\cite{DS2021Room}, further indicating the technological feasibility of VSD architecture.

\subsection{Security}

The VSD module introduces no additional security vulnerabilities. It operates exclusively on auxiliary DOFs that are fully independent of the message-encoding DOF and therefore neither leaks nor modifies the encoded quantum information, in principle.

From a theoretical perspective, an eavesdropper (Eve) is assumed to possess arbitrary physical capabilities. If exploiting the VSD process could provide any advantage, Eve could emulate an equivalent procedure within existing protocols, since VSD acts prior to receiver-side message decoding. From an information-theoretic standpoint, Eve’s optimal strategy is to maximize information gain under a fixed QBER, or equivalently, to minimize QBER for a given information gain. Any manipulation of the VSD module cannot reveal message information and can only increase the observed QBER, rendering such attacks counterproductive. Consequently, in the optimal scenario, Eve would prefer the receiver’s VSD module to operate ideally.

In principle, the security analysis of PRQC incorporating VSD is identical to that of standard PRQC, except that the effective NVSR is substantially enhanced. From a conceptual standpoint, VSD effectively re-normalizes the channel: it transforms a lossy channel with NVSR scaling as $O(t)$ into an effectively loss-insensitive one with NVSR scaling as $O(1)$ at the level of detection statistics, or equivalently, as enhancing the receiver’s ability to discriminate vacuum from non-vacuum events with high reliability. In either interpretation, Eve cannot achieve further benefits, and thereby existing security proofs remain applicable in principle. 

Indeed, as VSD is applied before the receiver's processing, PRQC with VSD can be viewed as traditional PRQC over an advanced channel that integrates ordinary channels with the VSD module. Since the security of traditional PRQC is required to be independent of channels that are assumed to be controlled by Eve, the secure proof of PRQC with VSD follows directly.

\subsection{Efficiency}

The efficiency of a standard PRQC protocol without VSD can be expressed as
\begin{equation}\begin{aligned}
R_{0} = S \times E(NVSR_{0}) \times g(QBER_{0}),
\end{aligned}\end{equation}
where $S$ is the sifting rate, $E(NVSR_{0})$ is the effective detection probability (monotonically increasing with $NVSR_{0}$), and $g(QBER_{0})$ denotes the post-processing efficiency (monotonically decreasing with $QBER_{0}$).

With the VSD module, the effective detection probability becomes
\begin{equation}\begin{aligned}
E_{\text{VSD}} = S_{\text{VSD}} \times E(NVSR),
\end{aligned}\end{equation}
where
\begin{equation}\begin{aligned}
S_{\text{VSD}} = tP_{\geq k|n} + (1-t)Q_{\geq k|n}
\end{aligned}\end{equation}
is the additional sifting factor introduced by VSD, and $P_{\geq k|n}$ ($Q_{\geq k|n}$) denotes the acceptance probability for non-empty (empty) signals. The overall efficiency is therefore given by
\begin{equation}\begin{aligned}
R = S \times S_{\text{VSD}} \times E(NVSR) \times g(QBER).
\end{aligned}\end{equation}

\begin{figure}[htbp]
  \centering
  \begin{subfigure}[htbp]{0.48\columnwidth}
    \centering
    \includegraphics[width=\linewidth]{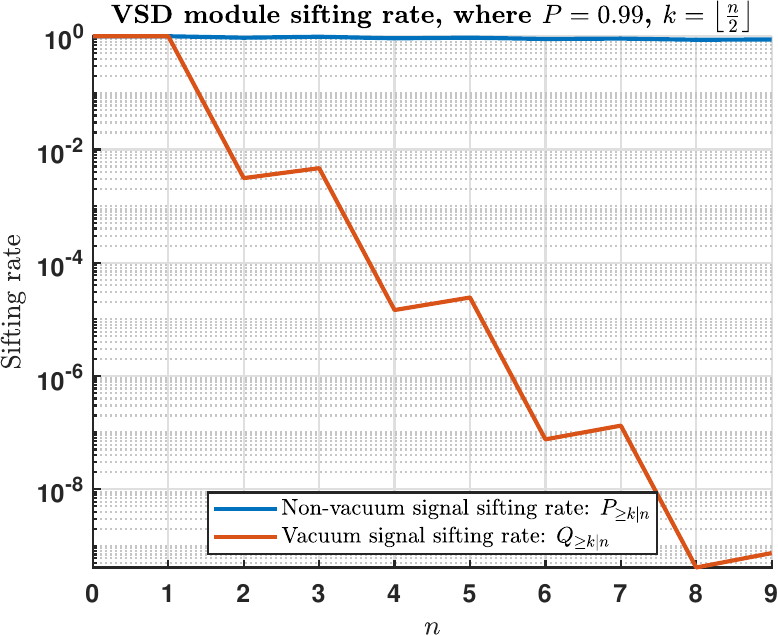}
    \caption{\textbf{Sifting rate vs $n$, $P=0.99$, $k=\left\lfloor\frac{n}{2}\right\rfloor$.}}
    \label{SVSD0.99}
  \end{subfigure}
  \hfill
  \begin{subfigure}[htbp]{0.48\columnwidth}
    \centering
    \includegraphics[width=\linewidth]{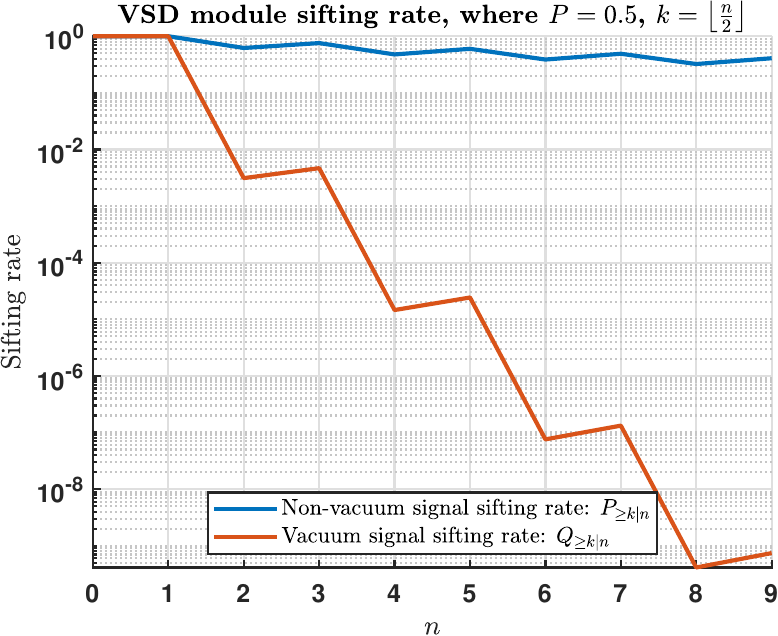}
    \caption{\textbf{Sifting rate vs $n$, $P=0.5$, $k=\left\lfloor\frac{n}{2}\right\rfloor$.}}
    \label{SVSD0.5}
  \end{subfigure}
  \hfill
  \begin{subfigure}[htbp]{0.48\columnwidth}
    \centering
    \includegraphics[width=\linewidth]{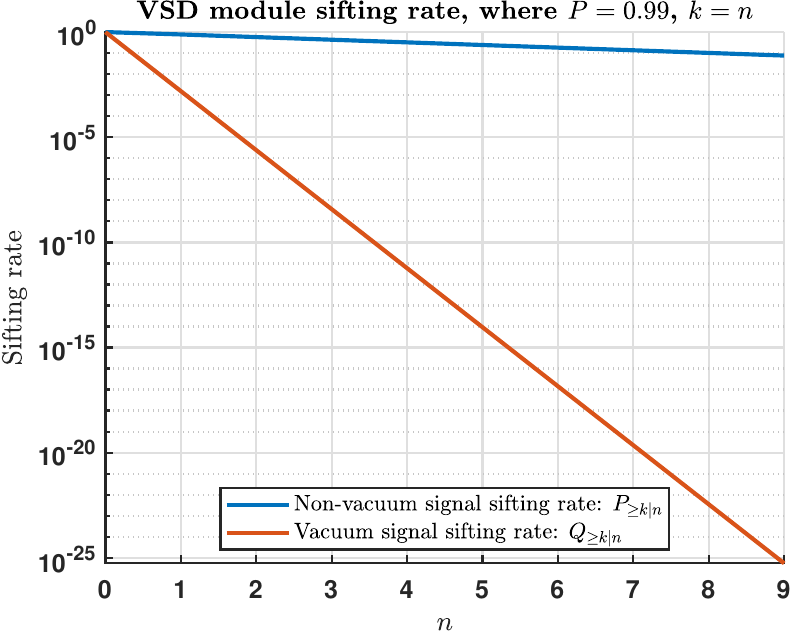}
    \caption{\textbf{Sifting rate vs $n$, $P=0.99$, $k=n$.}}
    \label{SVSDk=n0.99}
  \end{subfigure}
  \hfill
  \begin{subfigure}[htbp]{0.48\columnwidth}
    \centering
    \includegraphics[width=\linewidth]{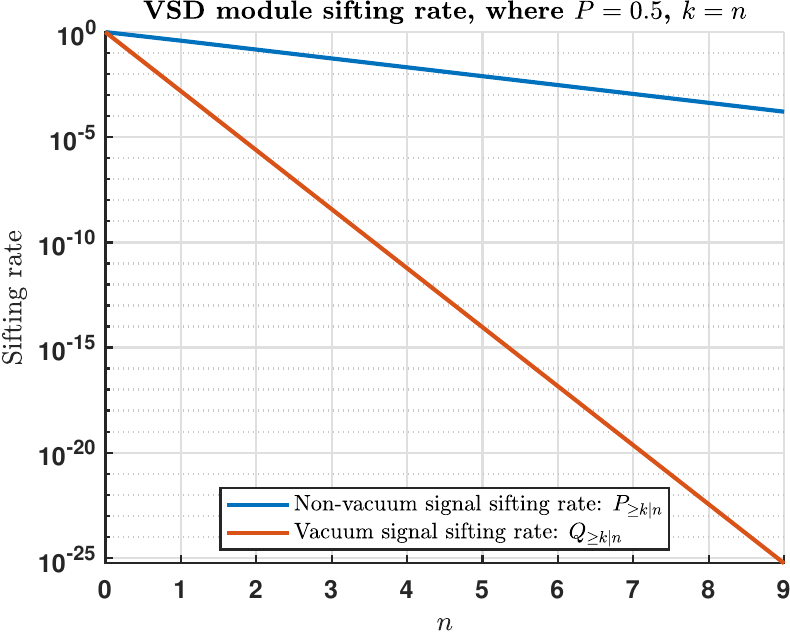}
    \caption{\textbf{Sifting rate vs $n$, $P=0.5$, $k=n$.}}
    \label{SVSDk=n0.5}
  \end{subfigure}
  \caption{\textbf{VSD sifting rate vs $n$ for different controlled-gate parameters.} Notations and settings follow the main text.}
  \label{Sifting}
\end{figure}

While the VSD framework can substantially enhance the NVSR and circumvent the QBER divergence, it introduces an intrinsic filtering overhead characterized by the sifting factor $S_{\text{VSD}}\leq1$. Although the discarded events predominantly originate from ineffective (vacuum) signals, a fraction of informative (non-vacuum) signals is inevitably suppressed, signifying a fundamental efficiency trade-off. Simulations of VSD sifting rates on vacuum signals and non-vacuum signals separately are presented in Fig. \ref{Sifting}, which indicates that with high-fidelity controlled gates (here with C-NOT), most discarded signals are vacuum, and only a few non-vacuum ones are discarded.

In the parameter space, increasing $n$ while reducing $k$ tends to optimize the sifting efficiency, whereas robust QBER suppression generally favors larger values for both $n$ and $k$. Furthermore, increasing $n$ entails a higher resource cost in terms of circuit complexity. Consequently, identifying the optimal operating point requires a synergistic balance between sifting efficiency, QBER suppression, and implementation overhead, even at the level of theoretical principle.

However, as demonstrated in Section~\ref{Experiment}, even modest VSD configurations can substantially extend the distance range over which PRQC remains operable, indicating that the trade-off between distance and efficiency can be quantitatively analyzed and optimized using numerical simulations.

In summary, although practical efficiency constraints remain, the VSD framework fundamentally alters the scaling limitation imposed by distance-dependent QBER, shifting PRQC from a regime limited by long-distance feasibility to one governed by efficiency considerations.

\subsection{Circuit Optimization}

\begin{figure}[htbp]
\centering
\includegraphics[width=0.9\linewidth]{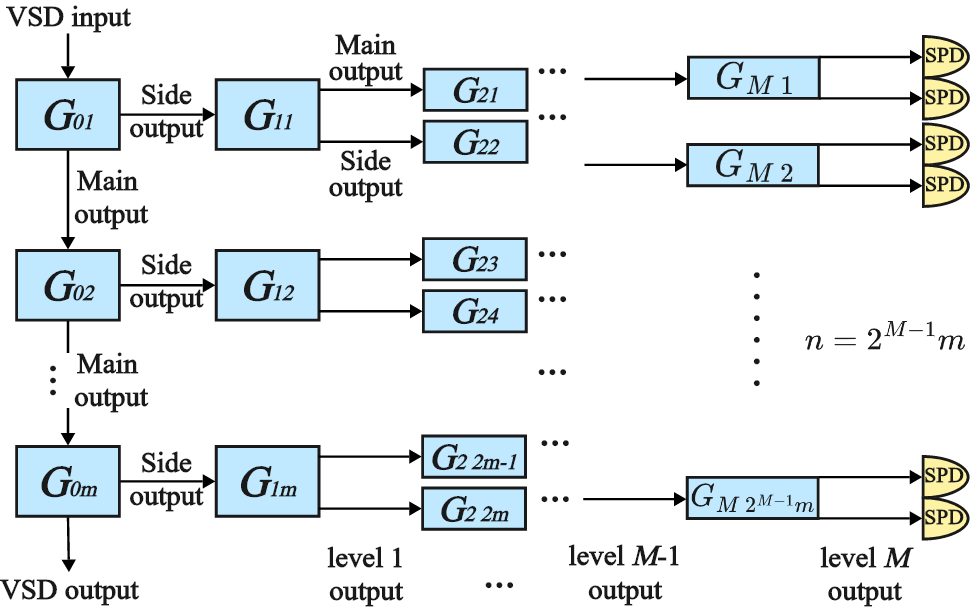}
\caption{\textbf{Potential parallelized architecture for the VSD module.}
Each block $G_{ij}$ is identical to the $G$ unit in Fig.~\ref{ESDM}, but arranged hierarchically to reduce circuit depth from $O(n)$ to $O(\log n)$.}
\label{IESDM}
\end{figure}

In quantum hardware design, reducing circuit depth is critical for mitigating cumulative errors. Since the VSD module comprises $n$ identical blocks, these operations may be amenable to parallel execution, enabling a significant reduction in circuit depth. Figure~\ref{IESDM} illustrates a potential alternative architecture in which the outputs of level $i>0$ serve as the inputs to level $i+1$, and only the outputs of level $M$ are finally detected. This hierarchical structure reduces the circuit depth from $O(n)$ to $O(\log n)$, offering a potentially more resource-efficient implementation. A detailed quantitative analysis of the performance and parameter trade-offs of this architecture is left for future work.

\subsection{Task Extension}

Although the VSD framework was initially developed to address the theoretical limitations of PRQC, the underlying principle of empty-signal discrimination may also be relevant to other quantum information tasks in which reliable identification of vacuum events is beneficial. For example, it could be applied in the conceptual design of SPDs, whose fundamental function is to distinguish vacuum from non-vacuum inputs, or in source engineering, where filtering vacuum components from weak coherent states is of interest for improving source quality. While such extensions lie beyond the scope of the present work and require further investigation, the VSD framework provides a general theoretical paradigm that may be applicable to a broader range of quantum technologies.

\section{Conclusion}

In this work, a rigorous theoretical framework termed empty-signal detection (VSD) has been introduced to address the long-standing problem of quantum bit error rate (QBER) explosion in practical repeaterless quantum communication (PRQC). From a result perspective, it provides a non-invasive mechanism to identify vacuum signals, which allows the non-vacuum signal ratio (NVSR) of the retained signals to be stabilized at a high value for arbitrary channel transmission rates, thereby suppressing the fundamental QBER to an acceptable level in PRQC at any distance. This demonstrates that the QBER divergence caused by channel attenuation and single-photon detector (SPD) dark counts, previously regarded as an intrinsic and insurmountable barrier to PRQC feasibility, can be formally circumvented at the theoretical level. Under mild and physically reasonable conditions, we prove that PRQC can remain viable, in principle, over arbitrarily long distances, which closes a conceptual gap in the theory of PRQC and clarifies its ultimate scalability. From a methodology perspective, the VSD method introduces a multi-copy analysis strategy for quantum communication, which was widely believed to be prevented by the no-cloning theorem. Its core is the decoupling of photon existence information (PEI) and the cryptographic message, where cloning the known PEI is not forbidden and can enhance the signal-to-noise ratio of quantum communication, similar to multi-copy analysis in classical communication.

More broadly, the VSD framework establishes a general, non-invasive, platform-agnostic principle for empty-signal discrimination. While formulated within the context of PRQC, the underlying logic of discriminating vacuum events through auxiliary coupling may find broader applications in various quantum tasks where loss plays a critical role. The framework presented here thus provides a conceptual foundation for future investigations into loss-resilient quantum communication and related quantum technologies.









\bibliographystyle{quantum}
\bibliography{QI}

\section{\label{Method}Appendix}

In this section, we rigorously establish the theoretical validity of the proposed VSD module. In particular, we present formal proofs of Theorem~\ref{Thm} and Corollary~\ref{Cor1}, demonstrating that the VSD module can suppress the fundamental QBER and, in principle, enable arbitrarily long-distance PRQC.

\subsection{Illustrations of the Controlled Operation}

A key component of the VSD module is the controlled operation $C(F)$. Importantly, $C(F)$ should be regarded as acting on three-dimensional (qutrit) Hilbert spaces rather than on two-dimensional (qubit) ones. The auxiliary DOF $L$ is represented by states $|x\rangle \in \mathbb{H}_{L}$ with orthonormal basis $\{|0\rangle, |1\rangle, |\emptyset\rangle\}$, where $|\emptyset\rangle$ denotes the vacuum state. For example, if $L$ corresponds to photon polarization, then $|0\rangle$ and $|1\rangle$ represent orthogonal polarization states of a non-empty signal, while $|\emptyset\rangle$ represents the vacuum.

The action of $C(F)$ is defined as follows:
\begin{itemize}
\item If the target is not initially in the state $|\emptyset\rangle$, it remains unchanged when the control is $|\emptyset\rangle$ or $|0\rangle$, and is transformed by $F$ when the control is $|1\rangle$.
\item If the target is initially $|\emptyset\rangle$, it remains unchanged.
\item The control state is preserved in all cases.
\item The operation is extended linearly over the entire Hilbert space.
\end{itemize}

This formulation ensures that $C(F)$ preserves the control particle, even in practical optical implementations where vacuum components are unavoidable.

\subsection{\label{NVSR}Calculations of NVSR}

We now derive the analytical expression for the NVSR and present the proof of Theorem~\ref{Thm}.

For $n$ auxiliary systems, the probability of obtaining at least $k$ detection events (i.e., clicks) for a non-empty input is given by
\begin{equation}\begin{aligned}
P_{\geq k|n}=&\sum_{i=1}^{n-1}p^{i-1}(1-p)\sum_{j_{1}+j_{2}\geq k}\binom{i}{j_{1}}P_{s}^{j_{1}}(1-P_{s})^{i-j_{1}}\binom{n-i}{j_{2}}Q_{s}^{j_{2}}(1-Q_{s})^{n-i-j_{2}} \\
&+p^{n-1}\sum_{i\geq k}\binom{n}{i}P_{s}^{i}(1-P_{s})^{n-i},
\end{aligned}\end{equation}
where the first term accounts for cases in which the control particle is lost before reaching the final controlled block, while the second term corresponds to cases where the input to the last block remains non-empty.

For a vacuum input, the probability of obtaining at least $k$ clicks is
\begin{equation}\begin{aligned}
Q_{\geq k|n}=\sum_{i\geq k}\binom{n}{i}Q_{s}^{i}(1-Q_{s})^{n-i}.
\end{aligned}\end{equation}

Furthermore, the probability of obtaining at least $k$ clicks while the VSD output remains non-empty is
\begin{equation}\begin{aligned}
P_{\geq k|n,VSD_{out}\neq\emptyset}=tp^{n}\sum_{i\geq k}\binom{n}{i}P_{s}^{i}(1-P_{s})^{n-i}.
\end{aligned}\end{equation}

Restricting attention to signals that pass the VSD check, the resulting NVSR is given by
\begin{equation}
\begin{aligned}
&NVSR=\frac{P_{\geq k|n,VSD_{out}\neq\emptyset}}{tP_{\geq k|n}+(1-t)Q_{\geq k|n}} \\ =&\frac{1}{\frac{1}{p}+\frac{1-p}{p^{n}}\frac{\sum_{i=1}^{n-1}p^{i-1}\sum_{j_{1}+j_{2}\geq k}\binom{i}{j_{1}}P_{s}^{j_{1}}(1-P_{s})^{i-j_{1}}\binom{n-i}{j_{2}}Q_{s}^{j_{2}}(1-Q_{s})^{n-i-j_{2}}}{\sum_{i\geq k}\binom{n}{i}P_{s}^{i}(1-P_{s})^{n-i}} +\frac{1-t}{tp^{n}}\frac{\sum_{i\geq k}\binom{n}{i}Q_{s}^{i}(1-Q_{s})^{n-i}}{\sum_{i\geq k}\binom{n}{i}P_{s}^{i}(1-P_{s})^{n-i}} }.
\end{aligned}
\end{equation}

Choosing $k=n$ simplifies the expression to
\begin{equation}
\begin{aligned}
NVSR|_{k=n}=\frac{1}{\frac{1}{p}+\frac{1-t}{t}\left(\frac{Q_{s}}{pP_{s}}\right)^{n}+\frac{1-p}{p}\sum_{i=1}^{n-1}\left(\frac{Q_{s}}{pP_{s}}\right)^{n-i}}.
\end{aligned}
\end{equation}

Taking the limit $n\to +\infty$ and using the condition $pP_{s}>Q_{s}$, we obtain
\begin{equation}
\begin{aligned}
\lim_{n\to +\infty}NVSR|_{k=n}
&=\lim_{n\to +\infty}\frac{1}{\frac{1}{p}+\frac{1-t}{t}\left(\frac{Q_{s}}{pP_{s}}\right)^{n}+\frac{1-p}{p}\sum_{i=1}^{n-1}\left(\frac{Q_{s}}{pP_{s}}\right)^{n-i}} \\
&=\frac{1}{\frac{1}{p}+\frac{1-p}{p}\frac{\frac{Q_{s}}{pP_{s}}}{1-\frac{Q_{s}}{pP_{s}}}}
=\frac{pP_{s}-Q_{s}}{P_{s}-Q_{s}}.
\end{aligned}
\end{equation}

Therefore, the NVSR specified in Theorem~\ref{Thm} can be asymptotically approached by choosing $k=n$ and increasing $n$\footnote{Indeed, for a fixed $n$, the choice $k=n$ yields the maximal NVSR, and increasing $n$ further enhances it.}. This completes the proof of Theorem~\ref{Thm}.

\subsection{\label{QBER}Calculations of QBER}

We now derive the expression for the QBER and prove Corollary~\ref{Cor1}.

For the message DOF $C$, let $c_{t}$ and $e_{t}$ denote the probabilities of correct and incorrect detections, respectively, when the processed signal is effective, and let $c_{l}$ and $e_{l}$ denote the corresponding probabilities when the signal is ineffective\footnote{These expressions assume two SPDs, where a valid detection event occurs if and only if exactly one SPD registers a click. For systems with more SPDs, $\eta_{1}$ and $d_{1}$ should represent the global detection efficiency and DCR, respectively.}.
These probabilities are given by
\begin{equation}\begin{aligned}
c_{t}=(1-e_{C})[\eta_{1}+(1-\eta_{1})d_{1}](1-d_{1})
+e_{C}(1-\eta_{1})d_{1}(1-d_{1}),
\end{aligned}\end{equation}
\begin{equation}\begin{aligned}
e_{t}=e_{C}[\eta_{1}+(1-\eta_{1})d_{1}](1-d_{1})
+(1-e_{C})(1-\eta_{1})d_{1}(1-d_{1}),
\end{aligned}\end{equation}
\begin{equation}\begin{aligned}
c_{l}=e_{l}=d_{1}(1-d_{1}).
\end{aligned}\end{equation}

The overall QBER after VSD filtering is therefore
\begin{equation}\begin{aligned}
\label{Q-N}
QBER=\frac{ESR\times e_{t}+(1-ESR)\times e_{l}}{ESR\times(e_{t}+c_{t})+(1-ESR)\times(e_{l}+c_{l})}.
\end{aligned}\end{equation}

\textbf{Proof of Proposition~\ref{QCThr} and Corollary~\ref{Cor1}:}

Since $e_{C}<e_{th}\leq 50\%$, it follows that
\begin{equation}
\begin{aligned}
&e_{th}>QBER\\
\Leftrightarrow\ 
& e_{th}>\frac{ESR\times e_{t}+(1-ESR)\times e_{l}}{ESR\times(e_{t}+c_{t})+(1-ESR)\times(e_{l}+c_{l})} \\
\Leftrightarrow\ 
& e_{th}(c_{t}+e_{t}-c_{l}-e_{l})ESR-(e_{t}-e_{l})ESR>e_{l}-e_{th}(c_{l}+e_{l}) \\
\Leftrightarrow\ 
& ESR>\frac{(1-2e_{th})d_{1}}{\eta_{1}[(e_{th}-e_{C})+d_{1}(1-2e_{th})]}:=\gamma.
\end{aligned}
\end{equation}
This completes the proof of Proposition~\ref{QCThr}.

According to Theorem~\ref{Thm}, under the condition of Corollary~\ref{Cor1}, the NVSR can asymptotically approach $\frac{pP_{s}-Q_{s}}{P_{s}-Q_{s}}$.
Hence,
\begin{equation}\begin{aligned}
\frac{pP_{s}-Q_{s}}{P_{s}-Q_{s}}>\gamma \ \Rightarrow\ \forall\ \text{channel transmission rate $t$},\ \text{QBER}<e_{th}\ \text{under suitable choice of $n,k$}.
\end{aligned}\end{equation}
since ESR$>$NVSR, and for the case that NVSR=ESR, the converse also holds. Expanding $P_{s}$ and $Q_{s}$ yields
\begin{equation}\begin{aligned}
\frac{pP_{s}-Q_{s}}{P_{s}-Q_{s}}>\gamma
\ \Leftrightarrow\ 
P(p-\gamma)>(1-p)\frac{d_{0}}{\eta_{0}(1-d_{0})}+(1-\gamma)Q.
\end{aligned}\end{equation}
This completes the proof of Corollary~\ref{Cor1}.
$\blacksquare$

\end{document}